\documentclass{article}
\usepackage{graphicx}
\usepackage{dcolumn}
\usepackage{bm}
\usepackage{hyperref}
\usepackage{subcaption}
\usepackage{fancyhdr}
\usepackage[utf8x]{inputenc} 
\usepackage{authblk}
\usepackage{array}
\title{Control of Graphene Layer Thickness Grown on Plasma Enhanced Atomic Layer Deposition of Molybdenum Carbide}
\author[1]{Eldad Grady}
\author[1]{W.M.M. Kessels}
\author[1]{Ageeth A. Bol}
\affil[1]{Department of Applied Physics, Eindhoven University of Technology, Den Dolech 2, P.O. Box 513, 5600 MB Eindhoven, The Netherlands}
\affil[ ]{\textit {Corresponding author: gradyel@protonmail.com}}

\date{}
\begin{document}
\maketitle
\begin{abstract}
 
We show the merits of plasma enhanced atomic layer deposition (PEALD) of catalytic substrate for chemical vapour deposition (CVD) graphene growth. The high quality multilayer graphene (MLG) on molybdenum carbide ($MoC_{x}$) thin film exhibits excellent uniformity and layer homogeneity over a large area. Moreover, we demonstrate how to achieve control of graphene layers thickness and properties, by varying the specific catalytic film chemical and physical properties. The control of growth is not digital, but is broad ranged from few layer graphene to a graphitic film of $\sim{75}$ graphene layers grown on the respective ALD catalytic substrates. Characterisation of the MLG has been performed using Raman spectroscopy, X-ray photoelectron spectroscopy (XPS), spectral ellipsometry (SE), and scanning low-energy electron microscopy (SLEEM).
By varying MLG thickness in a uniform homogeneous way, we can tailor the desired MLG properties for different application needs. Furthermore, the PEALD process can be readily adapted to high volume manufacturing processes, and combined with existing production lines. 

\end{abstract}
\fancyhead[]{Preprint}
\section{Introduction}
Graphene, a two dimensional sheet of carbon atoms, has attracted great interest in research because of its unique properties. A vast field of applications could utilise graphene to realise the next generation technology \cite{el2016graphene,bianco2018carbon}. Consequently, the ability to fabricate this material in a uniform scalable method and integrate into applications is the main challenge for both industry and academia. Various techniques have been explored to prepare graphene sheets, such as liquid phase exfoliation \cite{hernandez2008high}, epitaxy on SiC \cite{sutter2009epitaxial} and chemical vapour deposition (CVD) \cite{mattevi2011review}. Among these, CVD is the most promising for industrial scale production of high quality graphene. Graphene can be grown by CVD on a various of catalytic metallic substrates, including Cu, Co, Ni to name a few. CVD growth on Cu allows for growth of mostly single layer graphene sheets, but Cu grain size induce a polycrystalline graphene growth with defects on the graphene grain boundaries. Cu foils are commonly used for growth, and graphene defects form along the Cu railing sites. Sputtered Cu thin film can mitigate Cu foil related defects, but the thin film has its limitations as well \cite{lee2015suppression,tao2012synthesis}. Graphene growth is performed close to the melting point of Cu causing mass loss of Cu and subsequently holes in the film due to dewetting.
The difference between the thermal expansion in the Cu itself and the substrate cause for strain which leads to a wrinkly graphene sheet. Ni is commonly used for growth of multilayer graphene (MLG), due to the high solubility of carbon in Ni. While MLG grown on Ni is typically not uniform due to island like growth around the Ni grain boundaries, significant advancements have been made \cite{zakar2011nucleation, rameshan2018role}. \newline Recently, a scalable and uniform growth of MLG has been demonstrated on Mo thin films deposited by physical vapour deposition (PVD) \cite{GRACHOVA20141501,Ricciardella_2019}. Mo has the advantage of a higher melting point than Cu, which is vital to inhibit mass loss during growth. Furthermore, a similar thermal expansion coefficient to Si (4.8 and 2.6 $\mu{m}\cdot m^{-1}\cdot K^{-1}$ ) as oppose to Cu (17 $\mu{m}\cdot m^{-1}\cdot K^{-1}$), that reduces stress during growth and thus wrinkles in the graphene layers. Mo has also the advantage of being a clean room compatible material commonly used in IC manufacturing environment. However, PVD Mo thin films still suffer from from limitations related to the deposition technique. For the next generation applications, requirements for film uniformity, coverage and homogeneity are very strict, and PVD has reached its limitations in a high aspect ratio (HAR) topologies.
As PVD utilises highly kinetic ion bombardment to deposit the film, it is not suitable for coating MLG if a stack layer is required, as it will severely damage the MLG. Finally, impurities in sputtered Mo will also contribute to defects in the grown MLG. \newline
\linebreak
In order to answer the vast demands of graphene applications, such as high current density and specific capacitance, and to tailor them to specific application's needs, a control of graphene layers would be of high value. Extensive work on control of graphene layer numbers have been performed in recent years. Liu et. al. have shown that $H_2$ concentration and oxide nanoparticles influence the number of layers grown on Cu foil \cite{liu2015controllable}. However, the control was not uniform and varied within the graphene domain. Limbu et. al. \cite{limbu2018novel} have demonstrated layer control during hot fillament CVD growth on Cu (111) surface. However, the surface reaction growth is time dependent, and can take up to 200 mins to achieve few layer graphene. The first graphene adlayer is approximately 4 times longer than the first graphene layer, due to the lower adsorption energy of carbon on graphene than on Cu (111).\newline  
In this work we use atomic layer deposition techniques to fabricate the catalytic substrate for CVD graphene growth. Atomic layer deposition (ALD) allows for sub nanometre thickness control suitable for the strictest application demands to date. Additionally, we are able to tune the film composition and properties with high accuracy and reproducibility \cite{grady2019tailored}. By employing ALD we gain an additional merit of research, that allows us to study the effects and correlations between the two in a manner that is not possible with other deposition techniques mentioned above. As a result, we are able to present a controlled graphene layer thickness with a standard CVD growth process, by means of a bulk induced layer growth. The MLG grown were on molybdenum carbide ($MoC_{x}$) substrate that we fabricated by plasma enhanced atomic layer deposition (PEALD) \cite{grady2019tailored}. We refer to these samples as ALD $MoC_{x}$ films. Thus, high quality MLG were grown on large area, with excellent uniformity and homogeneity in CVD graphene fabrication. In the characterisation section we compare graphene grown on various types of $MoC_{x}$ films. An overview of results is presented, depicting the characterisation and a comparison of the grown MLG. The technique is scalable for large volume manufacturing and is compatible with IC and sensors fabrication environment.

\section{Experimental methods}
$MoC_{x}$ thin films have been deposited by plasma enhanced atomic layer deposition (PEALD)  at various temperatures and plasma conditions.
PEALD was performed on 100 mm Si (100) wafers coated with 450 nm of thermally grown $SiO_2$.  The depositions were performed in an Oxford instruments FlexAL2 ALD reactor, which is equipped with an inductively coupled remote RF plasma (ICP) source (13.56 MHz) with alumina dielectric tube. 
$MoC_{x}$ thin films have been deposited by PEALD  at various temperatures and plasma conditions, with $MoC_{x}$ films varying from 15$\mu{m}$ to 30$\mu{m}$ in thickness.
We define 4 generic types of $MoC_{x}$ films by their physical and chemical properties. Type A is defined as mostly amorphous film with low mass density and high carbon content (typically $C/Mo$ ratio $>0.9$). Type B has lower C/Mo ratio and higher mass density, amorphous film embedded with crystalline islands. Type C has high mass density ($\sim{9}$) low C/Mo ratio ($\sim{0.6}$) and is mostly amorphous. Type D films are highly crystalline cubic $\delta-MoC_{0.75}$, with high mass density. 
MLG was grown by low-pressure CVD (LPCVD) in a quartz tube (d=50mm, l=60cm) furnace with 3 heat zones set to 1100$^\circ$C. The typical base pressure when evacuated is $10^{-3}$ mbar. The furnace is set on cart wheels, to allow samples to be rapid annealed, as furnace temperature stabilises within 3.5 minutes after tube insertion. When moved away from the furnace, sample cooling down duration is typically 15 minutes.
Carbon feedstock gas ($CH_4$) is fed along with Argon through a quartz inner tube of 5 mm in diameter to the sealed side of the outer tube.
Graphene films  have been grown using CVD with $CH_4/Ar$ gas flow at 1100$^\circ$C after carbon saturation at a lower temperature. More details can be found elsewhere \cite{grady_eldad_2019_3542084}. 

\section{Characterisation of Graphene Films}
We studied the CVD grown MLG films, comparing graphene quality, uniformity and homogeneity.

Characterisation of the MLG has been performed using a Reinshaw InVia 514 nm laser Raman spectroscopy, X-ray photoelectron spectroscopy (XPS), spectral ellipsometry (SE) and scanning low-energy electron microscopy (SLEEM).
The thermal stability of $MoC_{x}$ ALD film was very high, and no delamination or mass loss was noticed for all measured film thicknesses. The experiments presented here have been performed on $\sim{25}\,\mu{m}$ thick $MoC_{x}$ films. A comparative analysis has been performed in order to understand the correlation between the catalytic substrate properties and the subsequent graphene layers grown on top.
Using PEALD $MoC_{x}$, we are able to control the characteristics $MoC_{x}$ film with excellent uniformity, by altering the $MoC_{x}$ film mass density and crystallinity as demonstrated in \cite{grady2019tailored}. 

\subsection{Raman Spectroscopy}

\begin{figure}[b]
	\centering
	\begin{subfigure}[t]{0.5\textwidth}
		\centering
		\includegraphics[width=\textwidth]{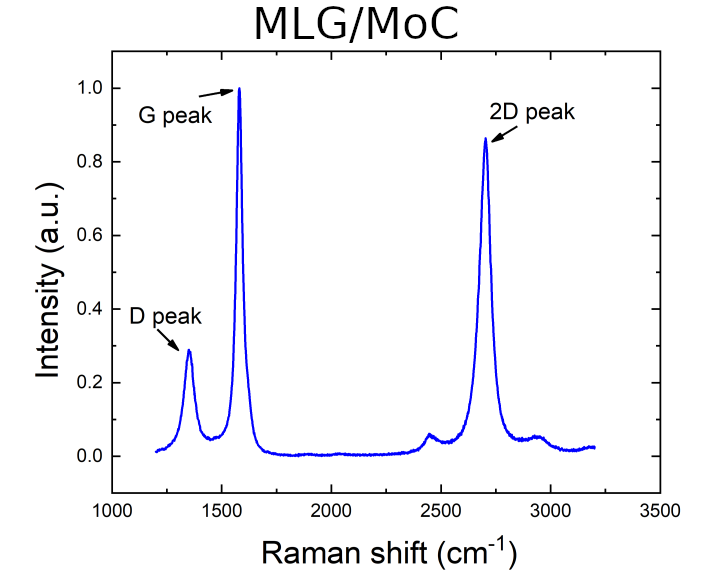}
		\caption{MLG/$MoC_{x}$ ALD}
		\label{fig:2D-G PVD}
	\end{subfigure}%
	\newline
	\begin{subfigure}[t]{0.5\textwidth}
		\centering
		\includegraphics[width=\textwidth]{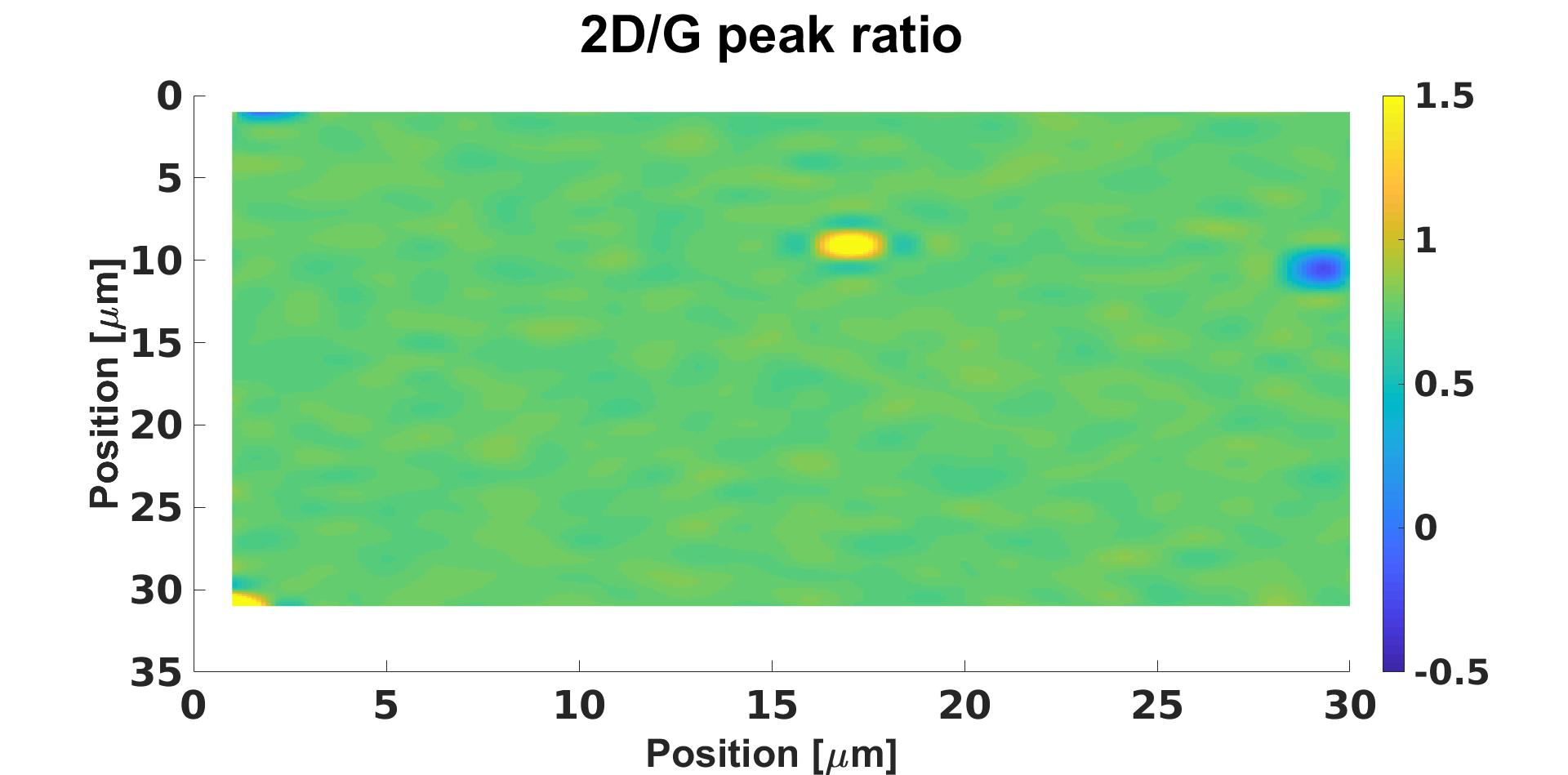}
		\caption{MLG/$MoC_{x}$ ALD}
		\label{fig:2D-G ALD}
	\end{subfigure}%
	\begin{subfigure}[t]{0.5\textwidth}
		\centering
		\includegraphics[width=\textwidth]{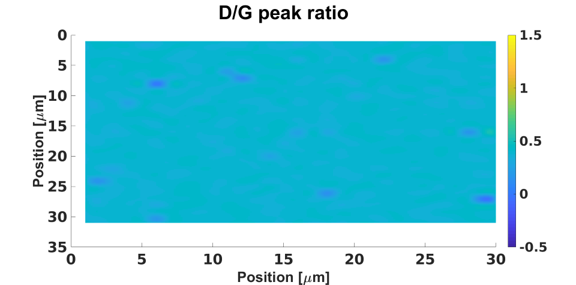}
		\caption{MLG/$MoC_{x}$ ALD}
		\label{fig:D-G ALD}
	\end{subfigure}%
	\caption{Raman spectra of MLG grown on 25 $\mu{m}$ thick $MoC_{x}$ ALD film. (a) Top: single Raman scan of the graphene. Bottom: Raman spectra of peaks ratio measured by a mapping scan of 30 × 30$\mu{m}^2$ area with 1$\mu{m}$ step resolution.   (b) Left: 2D/G peak ratio of MLG film. Variation in colour is inversely proportional to MLG uniformity. (c) Right: D/G peak ratio of MLG shows low D/G peak ratio with high uniformity.} 
	\label{fig:Raman map}
\end{figure}

Raman spectroscopy is a non-intrusive measurement which indicates graphene quality
by the characteristic bands of graphene at 1350 $cm^{-1}$ (D peak) 1580 $cm^{-1}$ (G peak) and
2700 $cm^{-1}$ (2D peak). In short, the D peak is related to disorder in the graphene lattice
and considered to indicate defects in the graphene layers. The ratio between 2D and G
peak indicates the nature of the graphene, with MLG ranges between 0.4-0.8 2D/G ratio,
while single layer graphene (SLG) ratio is typically $>1$. The FWHM for SLG is 30$cm^{-1}$
and for MLG ranges between 60 - 120 $cm^{-1}$ . Raman mapping scans were also used to characterise a large
area by scanning multiple adjacent spots. We have analysed graphene samples on a large scale with non-
intrusive methods, to gain statistical data on uniformity, homogeneity and defects of our
graphene samples. To the best of our knowledge, this is the first time graphene quality has
been quantified statistically by analysing single point Raman diagnostics of characteristic
graphene spectra to produce an overview of the entire scanned sheet attributes. We measured areas of 30x30 $\mu{m}^2$ with 1 $\mu{m}$ steps (laser spot resolution limit), to derive information on the MLG uniformity and defects variations.
To do so, the 2D/G and D/G ratios were calculated from $\sim$1000 scans for each area, for
statistical analysis. Variations in 2D/G are a good indicator for uniformity of MLG layers.
Both variation and value of D/G ratios are an indicator to amount of defects and uniformity
in MLG layers. We show the results both in form of colour maps to visualise uniformity
variations and as density distribution functions, to quantify these variations. The covariance
of each dataset indicates the uniformity and homogeneity of the graphene layers. As seen
in figure \ref{fig:2D-G ALD} the uniformity of MLG grown on $MoC_{x}$ ALD at 350$^{\circ}C$ is high, with typical ratio of MLG as well as FWHM of 2D peak. The narrow distribution function
shows high MLG uniformity with 2D/G ratio at about 0.9 and variance of 	$1.5\cdot10^{-2}$. Similarly, MLG grown on $MoC_{x}$ ALD film fabricated at low deposition temperature exhibits high uniformity and low defects ratio [see table \ref{fig:xps}].
As can be seen in figure \ref{fig:D-G ALD}, MLG grown on ALD film show lower average D/G peak ratios, indicating the overall high quality and uniformity, as confirmed by the low D/G covariance.

\subsection{X-ray photoelectron spectroscopy}
The film composition was analysed by X-ray photoelectron spectroscopy (XPS) with
a Thermo Scientific KA1066 spectrometer, using monochromatic Al $K\alpha$ x-rays with an
energy of 1486.6 eV. The films were sputtered with $Ar^{+}$ ion gun prior to scans, in order to
remove surface oxide and adventitious carbon. A continuous electron flood gun was employed
during measurements to compensate for charging. XPS is surface-sensitive
quantitative spectroscopic technique, that allows for analysis of a substrate surface chemical
composition. 
We compare MLG grown on various types of $MoC_{x}$ films with C1s core level spectra and reference it to graphene grown on Cu thin film. 

MLG grown on ALD films of type A shows pure carbon presence at the surface (within error margins) and no trace of the catalytic substrate or oxygen. The peak position confirms a $sp^{2}$ carbon bonds of MLG. As can be shown in figure \ref{fig:XPS_C1s_ALD}, the MLG has high purity $sp^2$ bonding with main contributing peak located at 284.4 eV. Due to the interaction of the photoelectron with free electrons in the film, a corresponding plasmon loss feature is
observed at $\sim{290.4}$eV signifying the graphene sheet conductivity. 
We can further examine the bulk of the material by etching with $Ar^+$ ion
gun to perform depth profiling and analyse the elemental composition and the nature of
the chemical bondings of each layer. A comparison of MLG grown on various types of $MoC_{x}$ ALD films is performed by etching 100 levels (2000 seconds). 
As can be shown in figure \ref{fig:XPS-controlofmlgthickness1}, the MLG varies in thickness between different type of catalytic substrates, as indicated by diminishing C1s peak intensities \cite{xu2010auger}. The diminished peak intensities are accompanied with the rise of Mo peak, which indicates a transition to the underlying catalytic substrate. The thinnest graphene sample was measured on type C $MoC_{x}$, with only 1 level with typical $sp^{2}$ carbon peak and $\sim{65} \,at.\%$ carbon on the surface, matching typical values measured for graphene grown on Cu thin film. The thickest MLG layers have shown a consistent $sp^{2}$ carbon peak for all 100 etched levels, with no indication of any other element peak. Carbon content remained stable around $\sim{98}\, at.\%$ as is shown in figure \ref*{XPS-Thick_film}. 

\subsection{Film Thickness Characterisation}
\subsubsection{ultra-low energy SEM/STEM of graphene}
The electron transmissivity at low energy (up to tens of eV) has proven a reliable tool for
counting graphene layers as an alternative to Raman spectroscopy, providing an enhanced
lateral resolution. Graphene layers exhibit contrasts connected with electron reflectivity
fluctuations below 8 eV and in an additional band around 15 eV in the ultra-low-energy
SEM. This phenomenon can also be employed for counting the graphene layers with a
correlation of n-1 minima reflectivity for n graphene layers. Studies of graphene layers
number and homogeneity were performed, as shown in figure \ref{fig:VASP} shows simulation of the fluctuations as a function of graphene layers as was simulated in VASP. Figure \ref{fig:number-of-layers} shows measured fluctuations on ALD $MoC_{x}$ MLG sample corresponding to 6 layers of graphene (5 minima). Measurements performed
concluded high layer homogeneity across the sample.
\subsubsection{Spectral ellipsometry}
 Film thickness and optical properties of the deposited films have been studied with a J.A.
Woollam UV-spectroscopic ellipsometer (SE). Data was obtained within the range of 190
nm 990 nm, and refractive index (n) and extinction coefficient (k) were determined. Figure
7 shows an extract of the layers thickness and composition, as modelled in CompleteEASE
software. Relevant selected oscillators were chosen to achieve optimal fitting of the data \cite{grady2019tailored}.\\
For a type A $MoC_{x}$ sample of $25\pm{1}$ nm thick film, we estimated an equivalent  $25\pm{1}$ thick MLG film grown on top. The MLG thickness corresponds to approximately 75 graphene layers, as was confirmed with XPS depth profiling matching all etch levels to $sp^{2}$ carbon peak positions.  
\section{Discussion and Conclusions}
We have demonstrated the effects of the catalytic substrate physical and chemical properties on the CVD grown MLG properties. The results presented have been collected over a time span of 18 months, with excellent reproducibility. While ALD reactor conditions may vary, the correlation between the ALD film and the CVD grown graphene remained identical.

With the ability to precisely control PEALD $MoC_{x}$ composition, we can control the grown MLG film ontop. $MoC_{x}$ film with low density, low crystallinity and high carbon content (type A) are readily able to saturate and precipitate free carbon content required to form $sp^{2}$ carbon bonds on the surface, as oppose to crystalline $MoC_{x}$ films with high mass density (type D). Moreover, these $MoC_{x}$ films will yield MLG with significantly lower defects.  $MoC_{x}$ mass density has shown to have a reverse proportion to MLG thickness, and a good correlation to defects ratio in MLG film, as Type C films have shown to inhibit graphene growth. 

The high crystalline type D film has shown to have high 2D/G ratio typical to FLG and a higher D/G ratio. The defects are likely occuring as the $MoC_{x}$ film transition from a cubic $MoC_{0.75}$ phase to orthorhombic phase during the high temperature MLG growth process \cite{grady_eldad_2019_3542084}.\newline
Type A films with the lowest mass density and highest C/Mo ratio have produced the thickest MLG layers. A film of $\sim{25}$ nm $MoC_{x}$ produced over 75 graphene layers, as was estimated by UV-SE. Type C high denisty film with low C/Mo ratio yielded the lowest number of MLG layers. We could accurately determine 6 graphene layers using SLEEM measurements grown on ALD $MoC_{x}$. Nevertheless, we stipulate thinner FLG could be achieved, as the Raman spectrum of type D graphene suggests. More work needs to be done in fine tuning MLG thickness by a study of the critical regimes of plasma time exposures, and more parameters should be looked at, specifically substrate thickness and crystallinity. In addition, other catalytic substrates should be evaluated, both as homogeneous elements and alloys, for suitability \cite{guo2018atomic}.

By profiling the chemical composition of 100 etch levels, we could confirm our stipulation of MLG thickness control, with proportional increase between graphene film thickness and initial $MoC_{x}$ film C/Mo ratio and inverse to mass density. This control can be achieved very accurately using PEALD film, which allows for reproducible and scalable fabrication. Furthermore, the catalytic substrate purity, homogeneity and uniformity plays an important role to the same properties of the graphene grown on top, namely MLG defects, homogeneity and uniformity.

Therefore, a uniform homogeneous film is vital in order to optimise MLG quality in a uniform manner. In this regard, ALD catalytic substrates would have an advantage over other commercial deposition techniques to achieve an overall higher quality film. In addition, the outstanding step coverage makes ALD a tool of choice if 3D objects are of interest.
By varying MLG thickness in a uniform homogeneous way, we can tailor the desired MLG properties for different application needs. Furthermore, the process is reproducible, and suitable for implementation in IC and sensors fabrication environments. The PEALD process can be readily adapted to high volume manufacturing processes, and combined with existing production lines. 

The MLG film grown on $MoC_{x}$ ALD film demonstrated here are of excellent quality and uniformity. The technique presented here has a significant advantage to other commercial CVD graphene fabrication techniques known thus far when scalability and compatibility to IC fabrication environment are the decisive factors.  The rapid advancement in ALD tools and techniques such as role to role, batch production, and selective area PEALD promise a leap forward in advancing commercial high quality graphene production. These can be combined to achieve graphene devices with atomic resolution and mitigate current edge placement errors in manufacturing by a complete bottom up process.
By utilising recent developments in the field of ALD, the path to next generation electronics based on high quality graphene is opened.

\subsection*{Acknowledgements}
This  research  is  supported  by  the Dutch  Technology  Foundation STW (project number 140930), which is part of the Netherlands Organization for Scientific Research (NWO), and partly funded by the Ministry of Economic Affairs as well as ASML and ZEISS. Eliska Mikmekova is acknowledged for SLEEM measurements and analysis of the data. E. Grady thanks Cristian Helvoirt, Janneke Zeegbregts, Jeroen van Gerwen and the lab technical staff for their support.
\bibliographystyle{unsrt}
\bibliography{MLG}

\begin{thebibliography}{10}

\bibitem{el2016graphene}
Maher~F El-Kady, Yuanlong Shao, and Richard~B Kaner.
\newblock Graphene for batteries, supercapacitors and beyond.
\newblock {\em Nature Reviews Materials}, 1(7):16033, 2016.

\bibitem{bianco2018carbon}
Alberto Bianco, Yongsheng Chen, Yuan Chen, Debjit Ghoshal, Robert~H Hurt,
  Yoong~Ahm Kim, Nikhil Koratkar, Vincent Meunier, and Mauricio Terrones.
\newblock A carbon science perspective in 2018: Current achievements and future
  challenges.
\newblock {\em Carbon}, 132:785--801, 2018.

\bibitem{hernandez2008high}
Yenny Hernandez, Valeria Nicolosi, Mustafa Lotya, Fiona~M Blighe, Zhenyu Sun,
  Sukanta De, IT~McGovern, Brendan Holland, Michele Byrne, Yurii~K Gun'Ko,
  et~al.
\newblock High-yield production of graphene by liquid-phase exfoliation of
  graphite.
\newblock {\em Nature nanotechnology}, 3(9):563, 2008.

\bibitem{sutter2009epitaxial}
Peter Sutter.
\newblock Epitaxial graphene: How silicon leaves the scene.
\newblock {\em Nature materials}, 8(3):171, 2009.

\bibitem{mattevi2011review}
Cecilia Mattevi, Hokwon Kim, and Manish Chhowalla.
\newblock A review of chemical vapour deposition of graphene on copper.
\newblock {\em Journal of Materials Chemistry}, 21(10):3324--3334, 2011.

\bibitem{lee2015suppression}
Alvin~L Lee, Li~Tao, and Deji Akinwande.
\newblock Suppression of copper thin film loss during graphene synthesis.
\newblock {\em ACS applied materials \& interfaces}, 7(3):1527--1532, 2015.

\bibitem{tao2012synthesis}
Li~Tao, Jongho Lee, Harry Chou, Milo Holt, Rodney~S Ruoff, and Deji Akinwande.
\newblock Synthesis of high quality monolayer graphene at reduced temperature
  on hydrogen-enriched evaporated copper (111) films.
\newblock {\em ACS nano}, 6(3):2319--2325, 2012.

\bibitem{zakar2011nucleation}
Eugene Zakar, Barbara~M Nichols, Stephen Kilpatrick, Gregory Meissner, Richard
  Fu, and Kevin Hauri.
\newblock Nucleation sites for multilayer graphene on nickel catalyst.
\newblock In {\em 2011 11th IEEE International Conference on Nanotechnology},
  pages 1516--1520. IEEE, 2011.

\bibitem{rameshan2018role}
Raffael Rameshan, Vedran Vonk, Dirk Franz, Jakub Drnec, Simon Penner, Andreas
  Garhofer, Florian Mittendorfer, Andreas Stierle, and Bernhard Kl{\"o}tzer.
\newblock Role of precursor carbides for graphene growth on ni (111).
\newblock {\em Scientific reports}, 8(1):2662, 2018.

\bibitem{GRACHOVA20141501}
Yelena Grachova, Sten Vollebregt, Andrea~Leonardo Lacaita, and Pasqualina~M.
  Sarro.
\newblock High quality wafer-scale cvd graphene on molybdenum thin film for
  sensing application.
\newblock {\em Procedia Engineering}, 87:1501 -- 1504, 2014.
\newblock EUROSENSORS 2014, the 28th European Conference on Solid-State
  Transducers.

\bibitem{Ricciardella_2019}
Filiberto Ricciardella, Sten Vollebregt, Evgenia Kurganova, A~J~M Giesbers,
  Majid Ahmadi, and Pasqualina~Maria Sarro.
\newblock Growth of multi-layered graphene on molybdenum catalyst by solid
  phase reaction with amorphous carbon.
\newblock {\em 2D Materials}, 6(3):035012, apr 2019.

\bibitem{liu2015controllable}
Jinyang Liu, Zhigao Huang, Fachun Lai, Limei Lin, Yangyang Xu, Chuandong Zuo,
  Weifeng Zheng, and Yan Qu.
\newblock Controllable growth of the graphene from millimeter-sized monolayer
  to multilayer on cu by chemical vapor deposition.
\newblock {\em Nanoscale research letters}, 10(1):455, 2015.

\bibitem{limbu2018novel}
Tej~B Limbu, Jean~C Hern\'andez, Frank Mendoza, Rajesh~K Katiyar, Joshua~James
  Razink, Vladimir~I Makarov, Brad~R Weiner, and Gerardo Morell.
\newblock A novel approach to the layer-number-controlled and
  grain-size-controlled growth of high quality graphene for nanoelectronics.
\newblock {\em ACS Applied Nano Materials}, 1(4):1502--1512, 2018.

\bibitem{grady2019tailored}
Eldad Grady, Marcel Verheijen, Tahsin Faraz, Saurabh Karwal, W.~M.~M. Kessels,
  and Ageeth~A. Bol.
\newblock Tailored molybdenum carbide properties and graphitic nano layer
  formation by plasma and ion energy control during plasma enhanced ald, 2019.

\bibitem{grady_eldad_2019_3542084}
Eldad Grady, Chenhui Li, Oded Raz, W.M.M. Kessels, and Ageeth~A. Bol.
\newblock {Resist and Transfer Free Patterned CVD Graphene Growth on ALD
  Molybdenum Carbide Nano Layers}.
\newblock {Preprint version}, November 2019.

\bibitem{xu2010auger}
Mingsheng Xu, Daisuke Fujita, Jianhua Gao, and Nobutaka Hanagata.
\newblock Auger electron spectroscopy: a rational method for determining
  thickness of graphene films.
\newblock {\em Acs Nano}, 4(5):2937--2945, 2010.

\bibitem{guo2018atomic}
Qun Guo, Zheng Guo, Jianmin Shi, Wei Xiong, Haibao Zhang, Qiang Chen, Zhongwei
  Liu, and Xinwei Wang.
\newblock Atomic layer deposition of nickel carbide from a nickel amidinate
  precursor and hydrogen plasma.
\newblock {\em ACS applied materials \& interfaces}, 10(9):8384--8390, 2018.

\end{thebibliography}

\newpage

\bigskip
\section{Figures}
\begin{table}[p]
	\begin{tabular}{|>{\centering}m {1.5cm}|>{\centering}m {1cm}|>{\centering}m {1cm}|>{\centering}m {1cm}|>{\centering}m {1cm}|>{\centering}m {1.5cm}|>{\centering}m {1cm}|>{\centering}m {1.5cm}|}
		\hline  
		\multicolumn{1}{|c|}{} & \multicolumn{3}{|c|} {XPS } & \multicolumn{2}{|c|} {Raman}  \tabularnewline
		\hline
		\rule[-2ex]{0pt}{5.5ex} Temp.($^{\circ}$C) &  [C] (at\%)& [O] (at\%)& [Mo] (at\%)   &
		2D/G& 2D/G variance & D/G& D/G variance  \tabularnewline 
		\hline
		\rule[-2ex]{0pt}{5.5ex} High temp. &  97.9 & $< d.l. $&  $< d.l. $  &0.96 &	$1.5\cdot10^{-2}$ & 0.19&  $7\cdot10{-4}$ \tabularnewline
		\rule[-2ex]{0pt}{5.5ex} Low temp. &  98.5 &  $< d.l. $ &  $< d.l. $ & 0.91& $2\cdot10^{-2}$ & 0.31& $1.2\cdot10^{-3}$  \tabularnewline

		\hline 
	\end{tabular} 
	\caption{MLG grown on type A $MoC_{x}$. Atomic \% of elements by XPS and summary of Raman scanning maps covariance values of the respective 2D/G and D/G peaks. A smaller covariance value corresponds to a higher uniformity MLG film. }
	\label{fig:xps}
\end{table}

\newpage

\begin{figure}[t!]
	\centering
	\begin{subfigure}[t]{0.5\textwidth}
		\centering
		\includegraphics[width=\textwidth]{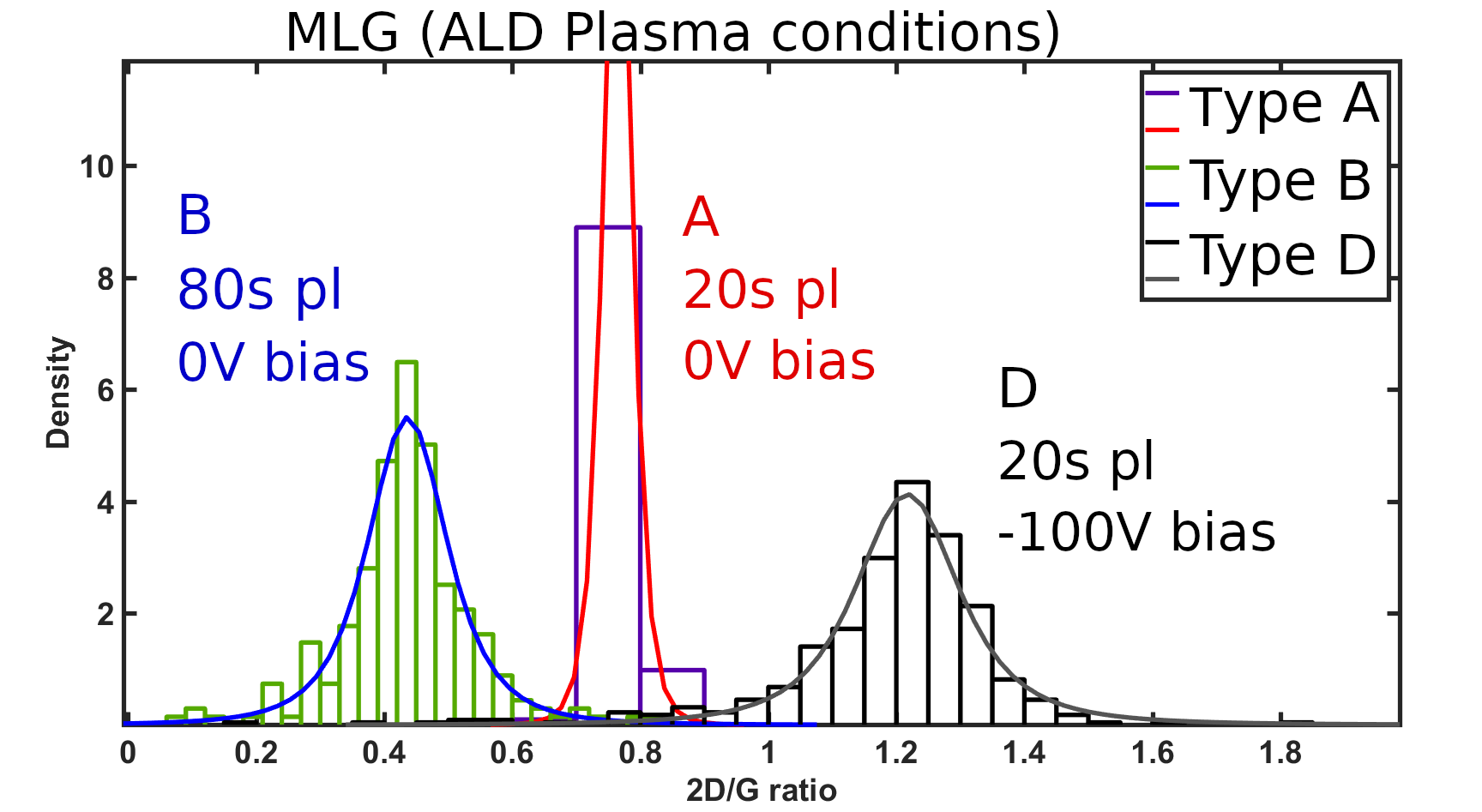}
		\caption{2D/G peak ratios distribution}
		\label{2D-G-distribution}
	\end{subfigure}%
	\begin{subfigure}[t]{0.5\textwidth}
		\centering
		\includegraphics[width=\textwidth]{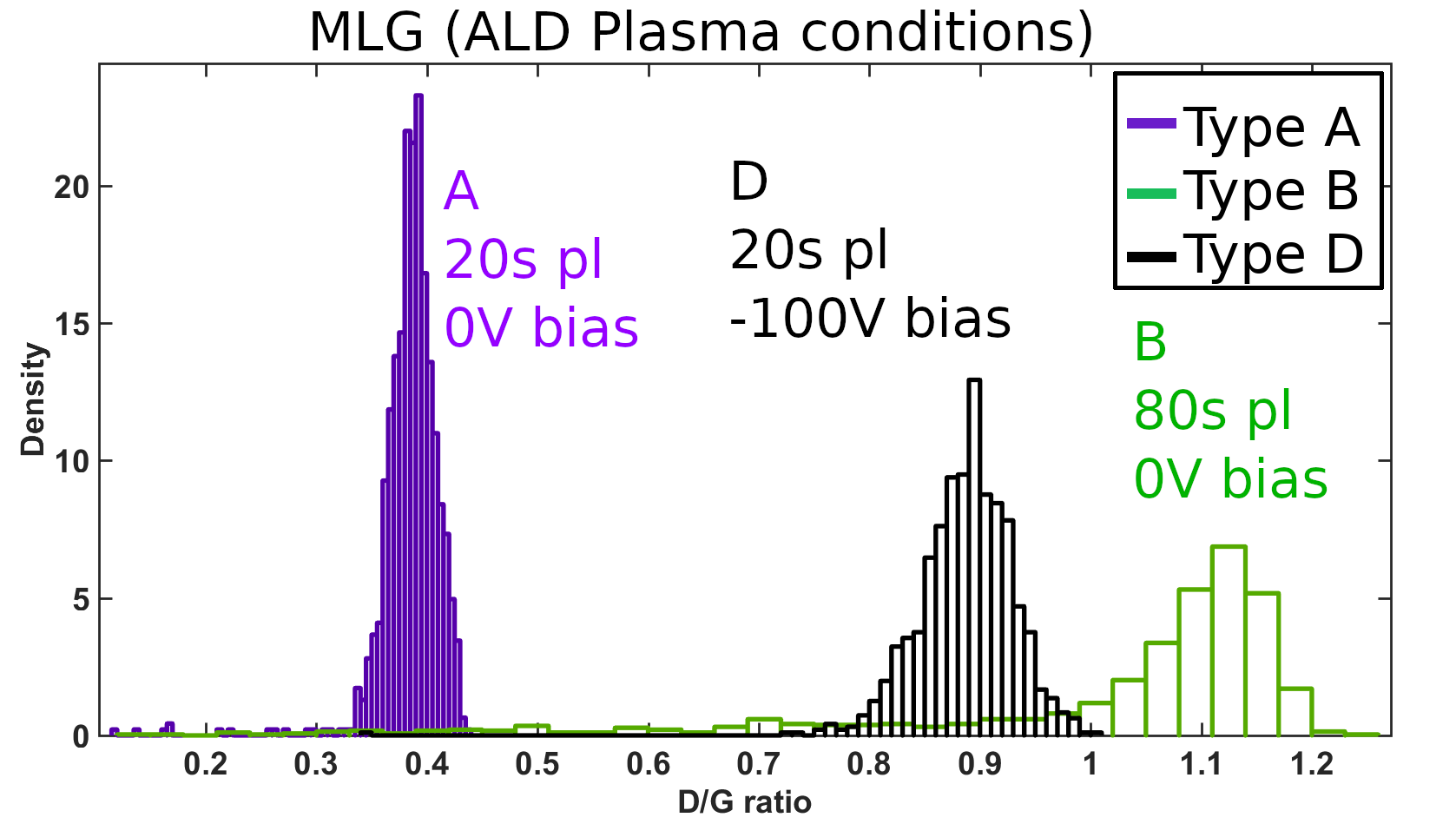}
		\caption{D/G peak ratios distribution}
		\label{D-G-distribution}
	\end{subfigure}%
	\caption{Distribution function of MLG Raman peak ratios, based on $\sim{1000}$ scanning points per function. Narrower distribution is generally attributed to higher uniformity of MLG. A narrower distribution of the 2D/G ratio indicates higher homogeneity of the MLG.}
	\label{fig:Ratio_distribution}
	
\end{figure}

\begin{figure}[t!]
	\centering

	\begin{subfigure}[t]{0.5\textwidth}
		\centering
		\includegraphics[width=\textwidth]{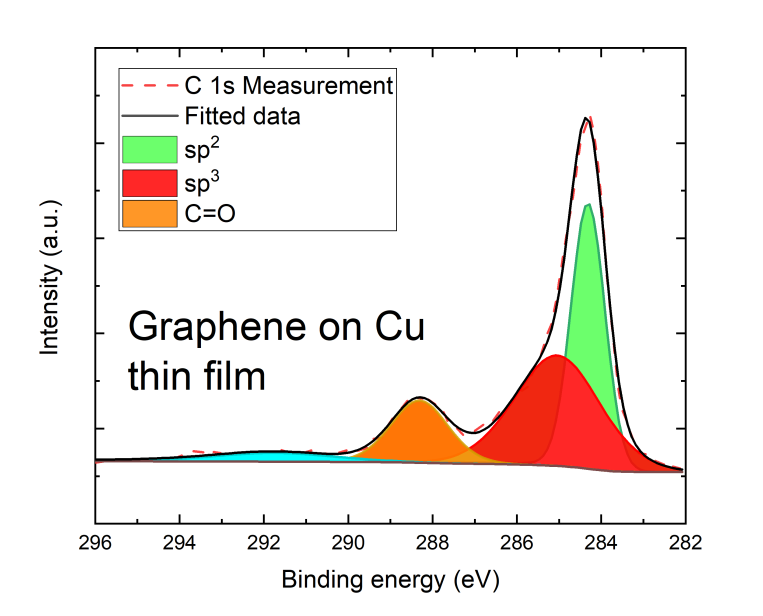}
		\caption{SLG/ Cu PVD C1s core level spectrum}
		\label{fig:XPS_C1s_Cu}
	\end{subfigure}%

	\begin{subfigure}[t]{0.5\textwidth}
		\centering
		\includegraphics[width=\textwidth]{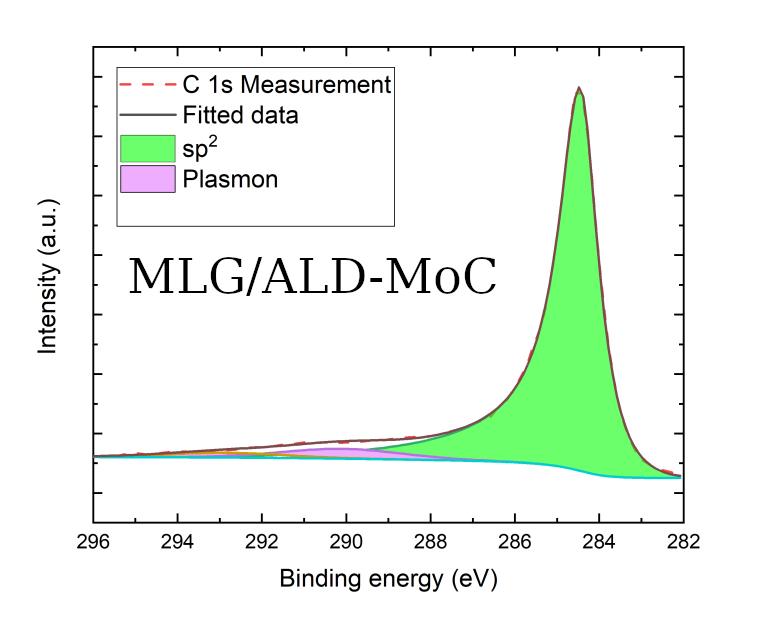}
		\caption{MLG/$MoC_{x}$ ALD C1s core level spectrum}
		\label{fig:XPS_C1s_ALD}
	\end{subfigure}%
	\caption{XPS spectra of C1s core level. (a) Left: XPS C 1s spectrum of single layer graphene (SLG) on Cu PVD thin film.(b) Right: C1s scan of MLG on ALD $MoC_{x}$. A typical asymmetric $sp^{2}$ peak at 284.4 eV indicate high purity graphene material. Plasmon loss features at $\sim{290.5}$ and  $\sim{293}$ are caused by interaction of the photoelectron with free electrons present in the graphene sheet. No sign of carbidic metal from catalytic substrate. }
	\label{fig:XPS_C1s}
\end{figure}

\begin{figure}[t!]
	\centering
	\begin{subfigure}[t]{0.5\textwidth}
		\centering
		\includegraphics[width=\textwidth]{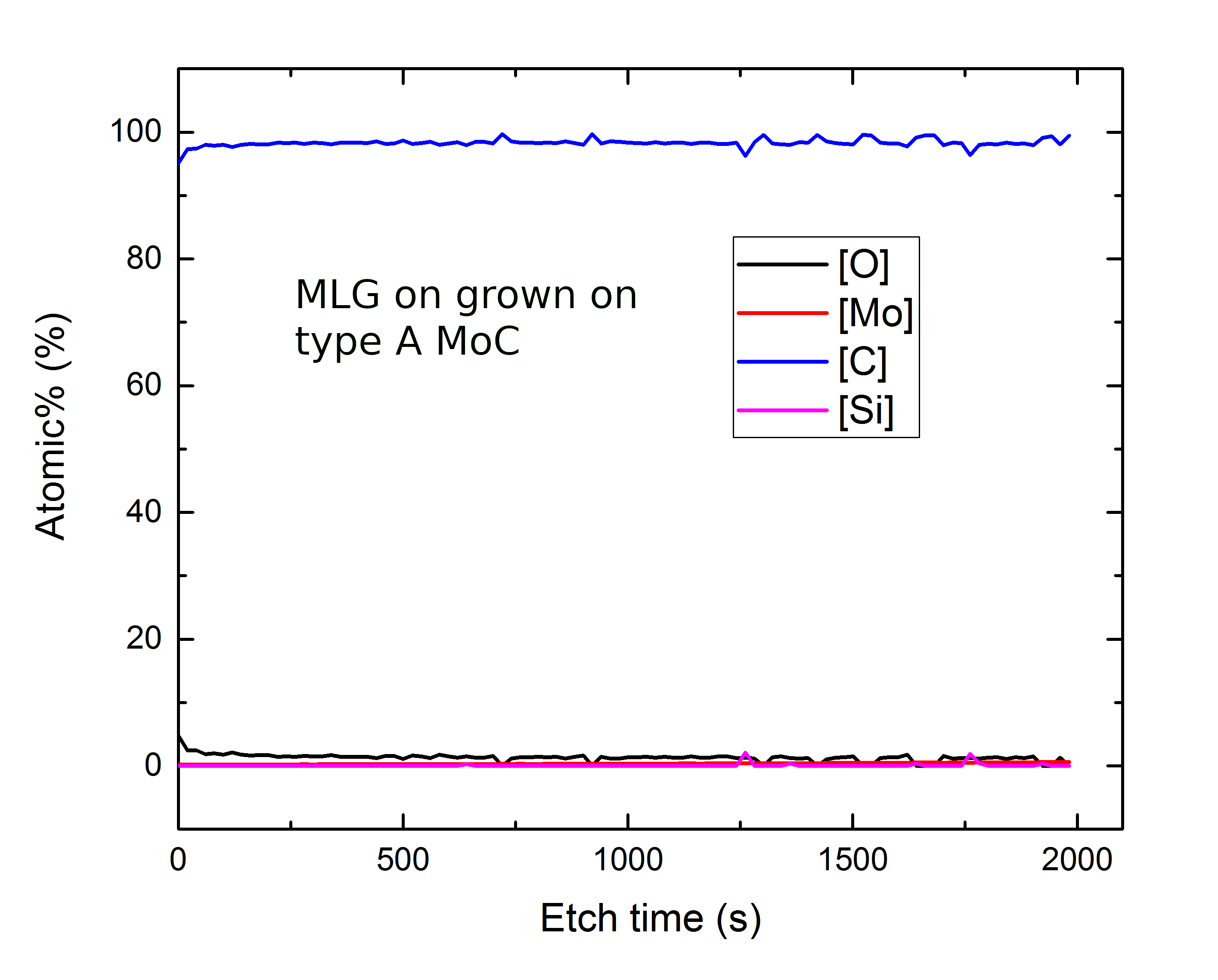}
		\caption{High purity graphene across 100 etch levels}
		\label{XPS-Thick_film}
	\end{subfigure}%
	\begin{subfigure}[t]{0.5\textwidth}
		\centering
		\includegraphics[width=\textwidth]{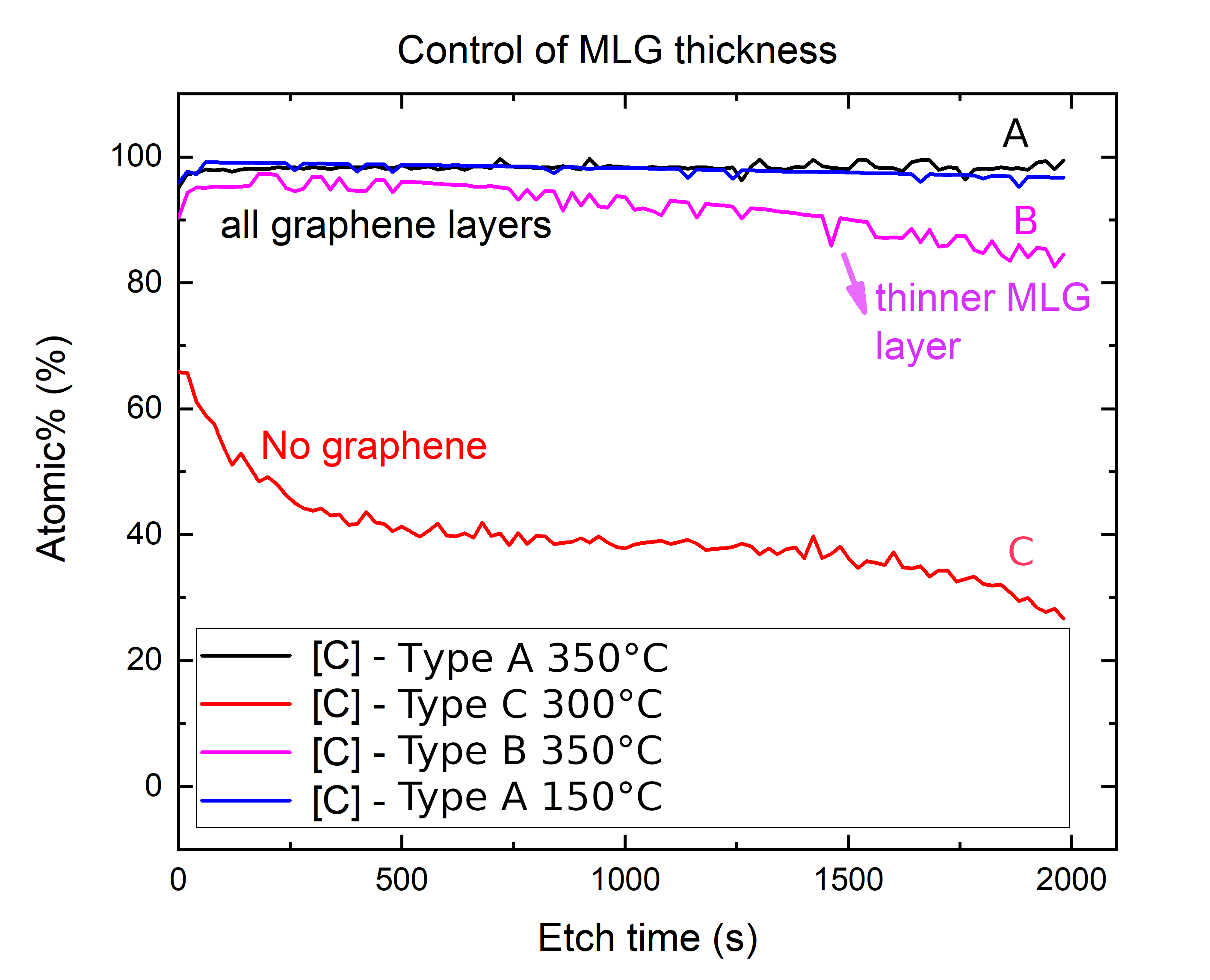}
		\caption{ Control of number of graphene layers}
	\end{subfigure}%
	\caption{XPS depth profiling of MLG on ALD catalytic substrate. (a) High purity $sp^{2}$ carbon content is shown in top figure for the entire 100 etch levels (20s etch time) for MLG grown on type A substrate. No impurities are measured within the XPS error range ($\pm{2}\,at.\%$). (b) C 1s spectra of samples deposited at various conditions, demonstrating the control of MLG layer thickness. Type C substrate inhibits MLG growth, with merely 65\% surface $sp^{2}$ carbon content, typical to $\leq{1}$ graphene layer.}
	\label{fig:XPS-controlofmlgthickness1}
	
\end{figure}

\begin{figure}
	\begin{subfigure}{0.9\textwidth}
		\centering
		\includegraphics[width=\textwidth]{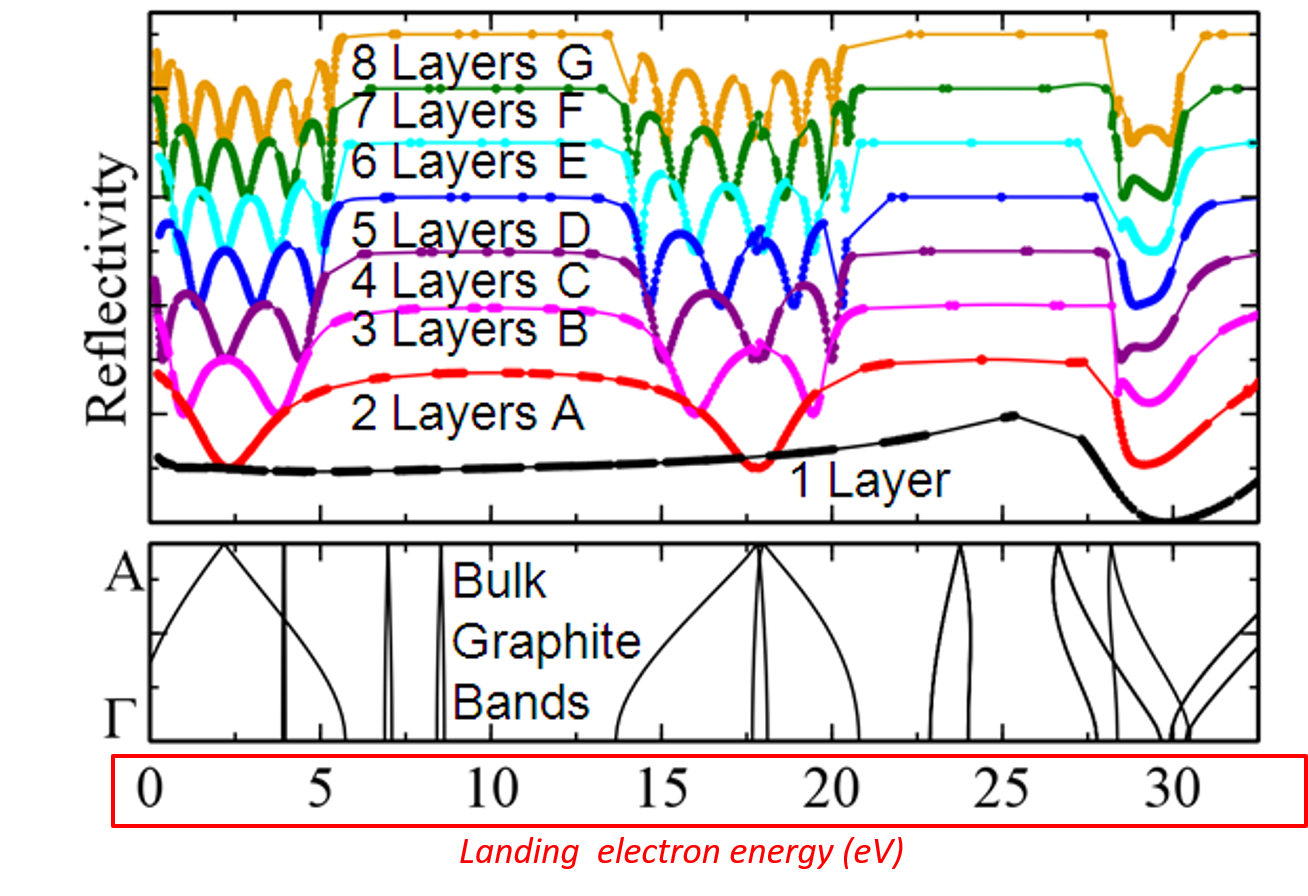}
		\caption{DFT simulations of electron reflectivity fluctuations as function of number
			of graphene layers}
		\label{fig:VASP}
	\end{subfigure}%
	\newline
	\begin{subfigure}{0.5\textwidth}
		\centering
		\includegraphics[width=\textwidth]{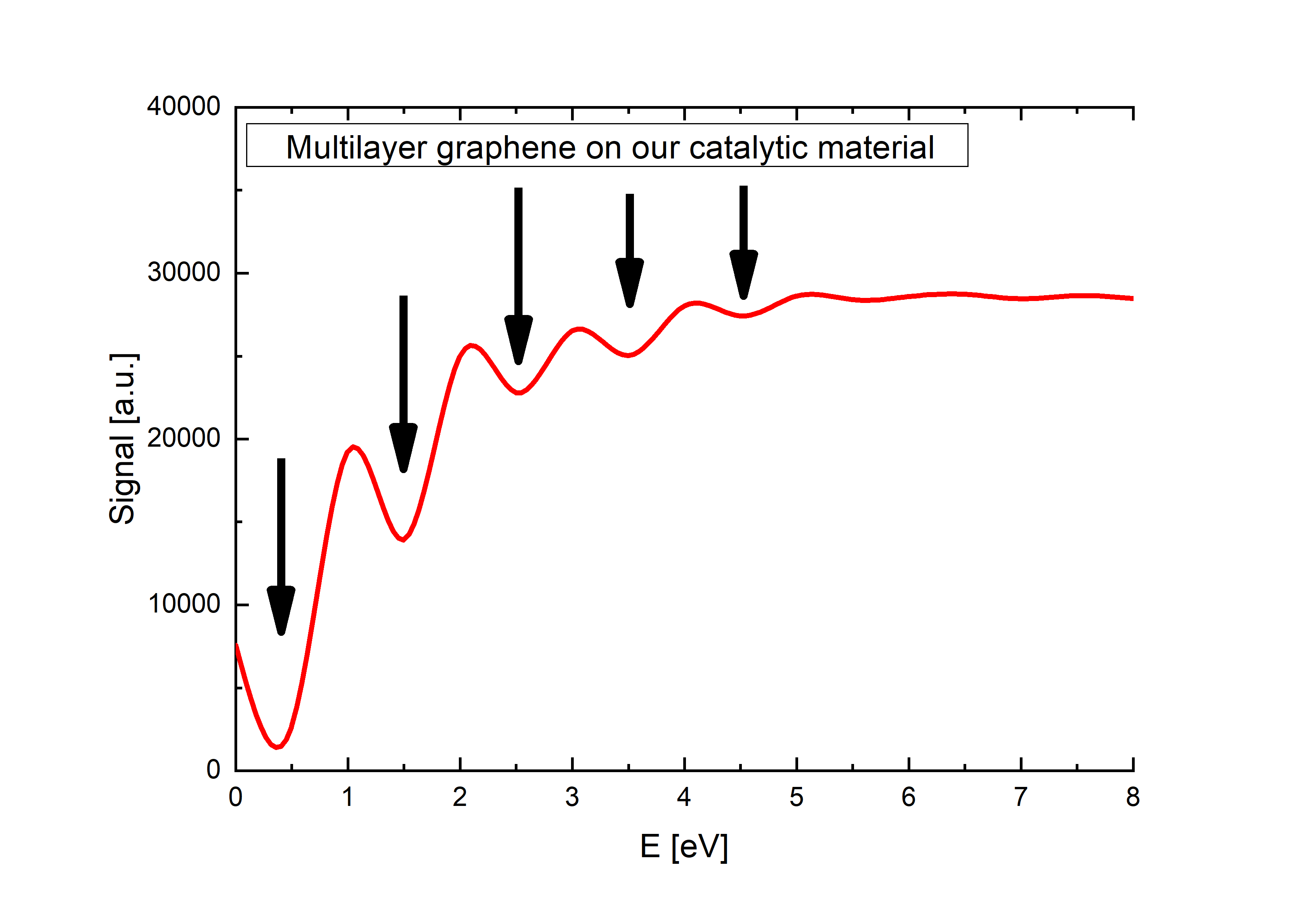}
		\caption{Lower energy bands}
		\label{fig:number-of-layers}
	\end{subfigure}%
	\begin{subfigure}{0.5\textwidth}
		\centering
		\includegraphics[width=\textwidth]{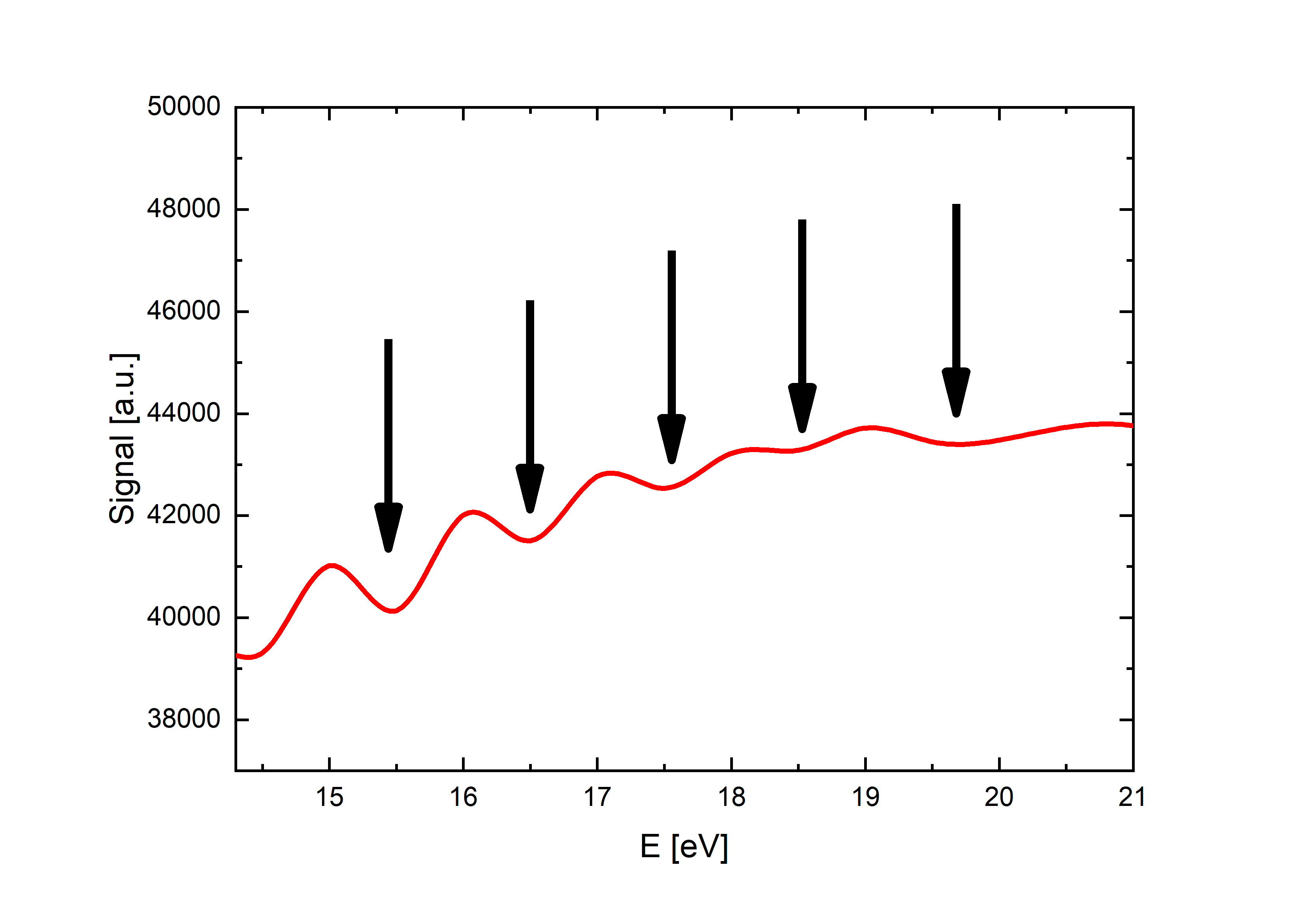}
		\caption{Higher energy bands}
	\end{subfigure}%
	\caption{Count of graphene layers by scanning low energy electron microscopy. Measurements of electron reflectivity in MLG/$MoC_{x}$ ALD sample at two energy bands. Simulations and measurements performed by Eliska Mikmekova (The Czech Academy of Sciences)}
	\label{fig:SLEEM}
	
\end{figure}

\begin{figure}
	\centering
	\includegraphics[width=0.9\linewidth]{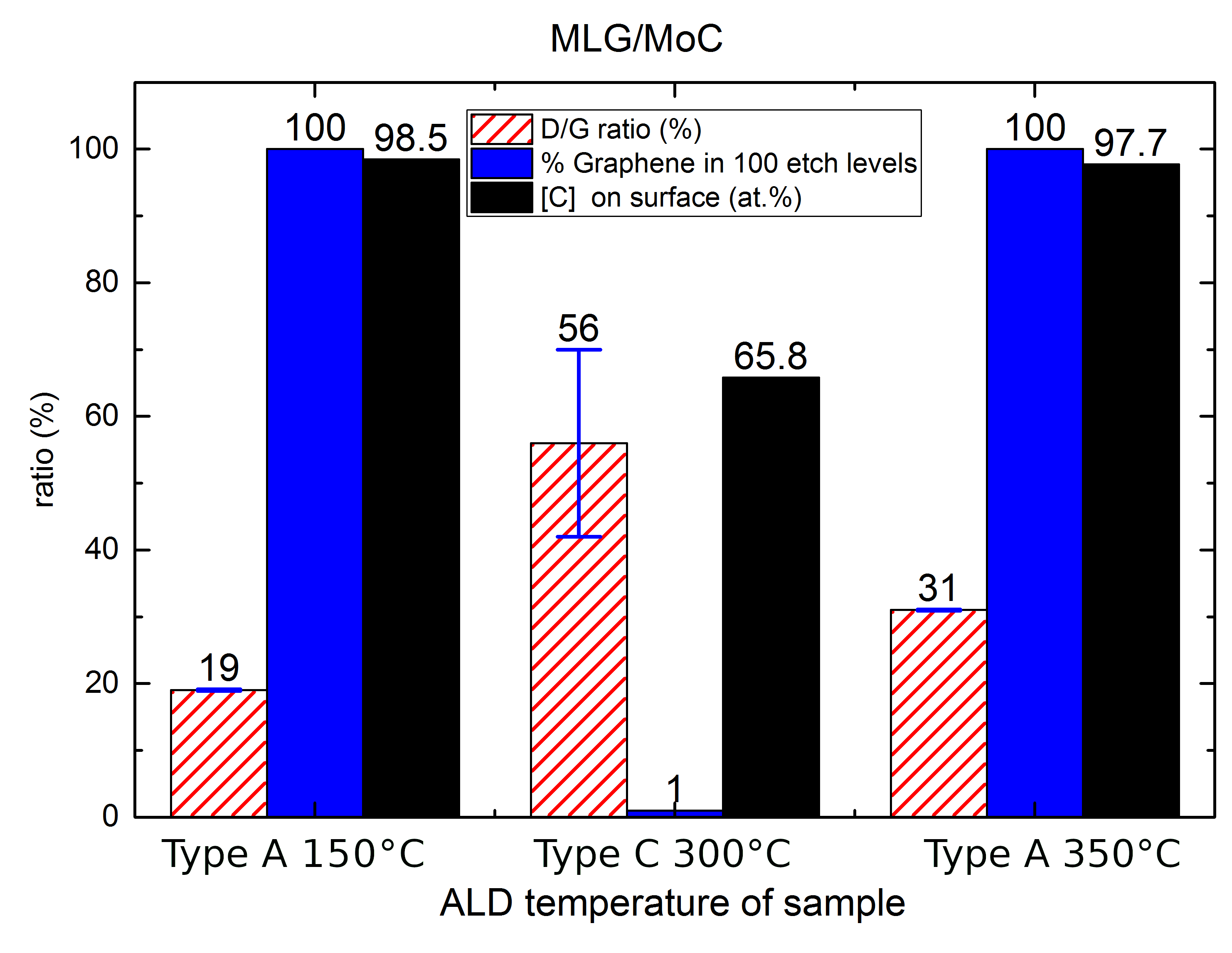}
	\caption{Overview of ALD $MoC_{x}$ temperature of deposition effects on graphene growth and properties. The effects of $MoC_{x}$ film on MLG are presented by comparison of D/G peaks as measurement with Raman mapping, \% of graphene as determined by $sp^{2}$ peak positions in each etch layer, and \% of carbon at the surface of the MLG that relates to the $sp^{2}$ peak position.}
	\label{fig:overview-ald}
\end{figure}

\end{document}